\documentclass[aps,pra,twocolumn, epsf]{revtex4}
\pdfoutput=1
\usepackage{graphicx}
\usepackage{amssymb}
\usepackage{amsmath}
\newcommand{\bra}[1]{\left\langle#1\right|}
\newcommand{\ket}[1]{\left|#1\right\rangle}
\newcommand{\adag}{{\hat{a}^\dag}}
\newcommand{\sigmdag}{{\hat {\sigma}^\dag}}
\newcommand{\sigm}{{\hat {\sigma}}}
\newcommand{\bdag}{{\hat{b}^\dag}}
\newcommand{\Hint}{\hat H_{\rm int}}
\newcommand{\piplus}{{\hat\pi}^{+}}
\newcommand{\piminus}{\hat\pi^-}
\newcommand{\piz}{\hat\pi_z}
\newcommand{\hc}{\mathrm{H.c.}}
\newcommand*{\pdiff}[3][]{\frac{\partial^{#1}{#2}}{\partial^{}{#3}{}^{#1}}}
\newcommand{\ave}[1]{\langle#1\rangle}
\newcommand{\eqnref}[1]{Eq.~(\ref{#1})} 
\newcommand{\St}{\hat {S}}
\newcommand{\Stdag}{\hat {S}^{\dag}}
\newcommand{\OmBog}{{\Omega_{\rm b}}}
\newcommand{\arctanh}{\textrm{arctanh}}
\newcommand{\abs}[1]{\left|#1\right|}
\newcommand{\hrho}{{\hat \rho}}
\newcommand{\rate}{r_{\rm at}}

\newcommand{\at}{{\rm at}}

\newcommand{\fref}{\ref}

\begin{document}

\title{Engineering atomic quantum reservoirs for photons}
\author{Susanne Pielawa$^{1,2}$, Luiz Davidovich$^3$, David Vitali$^4$, and Giovanna Morigi$^{1,5}$}

\affiliation{$^1$ Departament de F\'isica, Universitat Aut\`onoma de Barcelona, 08193 Bellaterra, Spain}
\affiliation{$^2$ Department of Physics, Harvard University, Cambridge MA 02138, U.S.A.}
\affiliation{$^3$ Instituto de Fisica, Universidade Federal do Rio de Janeiro, 21941-972 Rio de Janeiro, Brazil}
\affiliation{$^4$ Dipartimento di Fisica, Universit\`a di Camerino, 62032 Camerino, Italy}
\affiliation{$^5$ Theoretische Physik, Universit\"at des Saarlandes,  D-66041 Saarbr\"ucken, Germany}


\begin{abstract} We present protocols for creating entangled states of two modes of the electromagnetic field, by using a beam of atoms crossing microwave resonators. The atoms are driven by a transverse, classical field and pump correlated photons into (i) two modes of a cavity and (ii) the modes of two distant cavities. The protocols are based on a stochastic dynamics, characterized by random arrival times of the atoms and by random interaction times between atoms and cavity modes. The resulting effective model yields a master equation, whose steady state is an entangled state of the cavity modes. In this respect, the atoms act like a quantum reservoir, pulling the cavity modes into an entangled, Einstein-Podolski-Rosen (EPR) state, whose degree of entanglement is controlled by the intensity and the frequency of the transverse field. This scheme is robust against stochastic fluctuations in the atomic beam, and it does not require atomic detection nor velocity selection. 
\end{abstract}

\pacs{03.67.Bg, 03.65.Ud, 42.50.Pq, 42.50.Dv}

\date{\today} \maketitle

\section{Introduction}

The quest for quantum control of mesoscopic systems imposes the development of novel strategies, which go beyond the
implementation of fully coherent Hamiltonian dynamics. The main objective is to combat the effects of noise, which give rise to dissipation and decoherence. Protocols based on quantum error correction \cite{Gottesman, Knill}, quantum feedback \cite{Zippilli, Kubanek}, dynamical decoupling~\cite{Viola1, Viola2}, and decoherence-free subspaces~\cite{Zanardi, Lidar, Bacon} aim at minimizing the detrimental effects of coupling to an external environment. In this context, quantum reservoir engineering makes use of noise statistics as a resource for implementing robust quantum dynamics. The basic idea is to implement a stochastic dynamics whose stationary state is a nonclassical state. This is achieved by manipulating the coupling to the reservoir, whose properties are known only through the statistical averages~\cite{reservoir}. A prominent example of quantum reservoir engineering is laser cooling, which achieves low temperatures of single atoms or ions by tailoring the scattering cross section, such that in average the scattered photon carries away mechanical energy of the atomic center of mass~\cite{Metcalf,LaserCooling}.
More recently, this concept has been generalized in various directions, for the generation of nontrivial many-body states and nonequilibrium quantum phases
~\cite{PZnature}, and for the implementation of dissipation-driven quantum computation~\cite{ICnature}. 

The idea of using an atomic beam as a reservoir for a field in a cavity actually goes back to quantum laser theory, see for instance Ref.~\cite{scully-lamb}, where a thermal atomic beam crossing the laser cavity acts as a thermal bath for the laser field. Actual implementations of this idea have been made in experiments on cavity quantum electrodynamics, where a beam of atoms in the lower state of a two-level space is used to generate a vacuum state of the field in a microwave cavity, by absorbing thermal photons initially present in the cavity \cite{Englert:Lectures, master-eq, Nogues, QND-Nature}.

\begin{figure}[b]
\includegraphics[width=0.4\textwidth]{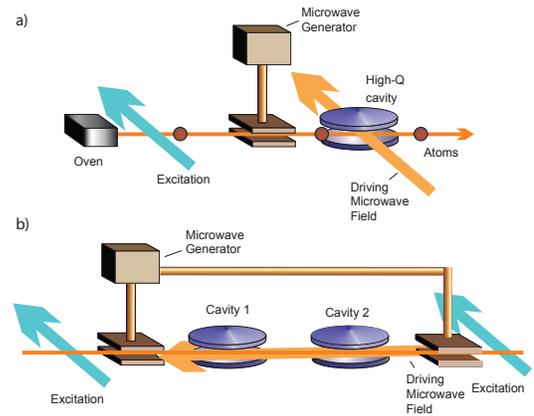} \caption{  
Setup of the system for creation of EPR state of  a)~two modes of the same cavity and b)~two spatially separated cavities. In both cases atoms from a beam are first prepared in a coherent
superposition of two Rydberg states $|g\rangle$ and $|e\rangle$ by
a combination of laser and microwave fields. The atoms have random
arrival times, and a low pumping rate warrants that at most one
atom is inside the resonator at a time~\cite{Haroche-Colloquium,
CavityQED-Walther}. While in the cavities, the dipole transition
$\ket g \rightarrow \ket e$ is saturated by a transverse microwave
field, thereby pumping on resonance two nondegenerate modes of either one or two cavites,  which are led asymptotically to a two-mode
squeezed state.} \label{Fig:exp-setup} \end{figure}

Recently, we proposed a method for preparing quantum states of the
electromagnetic field based on quantum reservoir
engineering~\cite{Pielawa07}. This method is implemented in a
typical setup of microwave Cavity Quantum Electrodynamics (CQED) as in
Fig.~\ref{Fig:exp-setup}(a), where the resonator is pumped by a beam
of atoms with random arrival times, and needs neither atomic
detection, nor detailed control of the sequence of atoms. We
showed that, by suitably preparing the initial state of the
incoming atoms, two-mode squeezing, i.e., Einstein-Podolski-Rosen
correlations~\cite{EPR}, is created between the cavity modes at
steady state. In this paper we provide a detailed discussion of
the proposal and analyze its robustness. We show, moreover, that
further nonclassical states of the
electromagnetic field can be realized by tuning different parameters of the setup in Fig.~\ref{Fig:exp-setup}(a), and which are stationary states of the
interaction with the atomic beam. 

We also propose a scheme for
entangling two distant resonators in a setup like the one sketched in Fig.~\ref{Fig:exp-setup}(b). Our procedure extends to the preparation of nonlocal EPR states previous ideas regarding the production of macroscopically separated fields \cite{scully-walther,meystre,davidovich, PRA603229}.

This article is organized as follows. In Sec.~\ref{Sec:2} we
review the basic concepts at the basis of quantum reservoir
engineering in microwave cavity QED. In Sec.~\ref{Sec:3} we
discuss the specific scheme for creating EPR states of the modes
of a resonator. In Sec.~\ref{Sec:4} a method for
entangling the modes of two distant cavities is proposed. The conclusions are
presented in Sec.~\ref{Sec:5}.

\section{An atomic reservoir for microwave photons} \label{Sec:2}

A typical setup of microwave cavity QED is sketched in Fig.~\ref{Fig:exp-setup}(a).
Preparation and monitoring of the cavity field is achieved by interaction with atomic beams,
whose internal state is prepared in a circular Rydberg state with large principal number, and
possessing a dipolar transition which couples resonantly with the
cavity mode. While the internal state of the atoms and their
velocity can be prepared with high precision, the atomic arrival
time is a stochastic variable which is known only
probabilistically, according to a Poissonian distribution. The
interaction of the cavity field with individual atoms is hence
warranted by setting very low rates, such that the probability
that two atoms are found inside the resonator is
negligible~\cite{CavityQED-Walther,Haroche-Colloquium}. The interaction between atoms and cavity mode is coherent to a large extent:
The atomic circular Rydberg states are typically stable over the interaction time, and high-finesse resonators in microwave cavity QED can reach very
long photon-storage times, which in state-of-the-art experiments can reach the order of fractions of seconds~\cite{CavityQED-Walther,ultra}.
These properties allowed for the realization of milestone experiments. Some
paradigmatic ones are the preparation and measurement of
nonclassical states of the microwave
field~\cite{Haroche-Colloquium,CavityQED-Walther}, the
experimental characterization of loss of coherence of the quantum
field~\cite{Decoherence} and of the transition from quantum to
classical dynamics~\cite{Quantum-Classical}, and the quantum
nondemolition measurement of the number of photons of the cavity
field~\cite{QND-Nature, Nature448-889}.

Most of these formidable results were obtained implementing Hamiltonian dynamics, where the interaction between atoms and cavity photons is essentially dispersive.
On the contrary, the realization of nonclassical states of the cavity field in the
dissipative regime, where atoms and photons exchange energy, is
based on properly tailored dissipative dynamics, where the atoms
act as reservoir of the photonic field.  As mentioned before, a simple example is the
preparation of the cavity mode in the vacuum state~ \cite{Englert:Lectures, master-eq, Nogues, QND-Nature}. Creation of
other Fock states of the cavity field has been achieved in
milestones experiments made in the strong-coupling regime, when
the resonator field saturates the atomic transition~\cite{WaltherTrapping,WaltherNature}. In this case,
by accurately selecting the atomic velocity so that the
interaction time of each atom is fixed, the nonlinear dynamics of
atom and cavity field possesses several fixed points, so-called
trapping states~\cite{Meystre}, which approach Fock states of the cavity field in the limit of negligible dissipation.

Several theoretical proposals have been discussed in the
literature, which provide schemes for the preparation of an arbitrary single-mode quantum
state of the electromagnetic field in a resonator, involving
resonant interaction with a well-controlled sequence of atoms,
without the need of atomic detection~\cite{Zoller,Law,Wellens00}.
This latter requirement is indeed important, since present
experiments lack high-efficiency detectors.

In this section we briefly review the basic properties of atom-photon interactions in microwave CQED from the point of
view of quantum reservoir engineering, hence setting the ground for the proposals for establishing EPR-correlations presented in
Sec.~\ref{Sec:3} and Sec.~\ref{Sec:4} of this article. For a comprehensive review the reader is referred to Refs.~\cite{Englert:Lectures,CavityQED-Walther,Haroche-Colloquium,HarocheBook}

\subsection{Jaynes-Cummings Hamiltonian}

The stability of the atomic states and of the resonator mode
during the interaction time justifies the use of Hamiltonian
dynamics for the interaction between a single atom and the cavity
mode. The Hamiltonian governing the dynamics of a single dipole
and the cavity mode is well described by the Jaynes-Cummings
Hamiltonian, $\hat H=\hat H_0+\hat H_{\rm JC}$. Here
$\hat H_0=\hbar\omega_0\hat \sigma^{\dagger}\hat \sigma
+\hbar\omega_c\hat a^{\dagger}\hat a$ gives the unperturbed evolution of a
dipole at frequency $\omega_0$ and a harmonic oscillator, the
cavity mode, at frequency $\omega_c$, while the interaction
between the dipole and the electric field of the cavity reads
\begin{equation}\label{Eq:JC} \hat H_{\rm JC}=\hbar g \hat a^{\dagger}\hat \sigma
+{\rm H.c.}, \end{equation} with $g$ the coupling strength. Here,
$\hat a$, $\hat a^{\dagger}$ are the annihilation, creation operator of a
cavity photon, while $\hat \sigma=|g\rangle\langle e|$,
$\hat \sigma^{\dagger}=|e\rangle\langle g|$ describe the dipole
lowering and rising operators, with $|g\rangle$ and $|e\rangle$
the ground and excited states, respectively, of the dipolar
transition.

For $\omega_0=\omega_c$, then $[\hat H_0,\hat H_{\rm JC}]=0$ and the
evolution operator can be written as \begin{equation}
\hat U(t)=\exp\left(-{\rm i}\hat Ht/\hbar\right)=\exp\left(-{\rm
i}\hat H_0t/\hbar\right)\exp\left(-{\rm i}\hat H_{\rm JC}t/\hbar\right)
\end{equation} whereby \begin{eqnarray} &&{\rm e}^{-{\rm i}\hat H_{\rm
JC}t/\hbar}|g\rangle=\cos(\phi \sqrt{\adag \hat a})|g\rangle+{\rm
i}\frac{\sin(\phi\sqrt{\hat a\adag})}{\sqrt{\hat
a\adag}}\hat a|e\rangle\label{U:1}\\
&&{\rm e}^{-{\rm i}\hat H_{\rm JC}t/\hbar}|e\rangle=\cos(\phi \sqrt{
\hat a\adag})|e\rangle+{\rm i}\adag\frac{\sin(\phi\sqrt{\hat
a\adag})}{\sqrt{\hat a\adag}}|g\rangle \label{U:2}\end{eqnarray}
and $\phi=g t$ is the Rabi angle. These equations show explicitly
the periodic exchange of energy between dipole and field when they
are resonantly coupled.

In the rest of this section we will assume that cavity mode and
atomic dipole are resonant, $\omega_c=\omega_0$.

\subsection{Interaction with a beam of atoms} \label{Sec:2:0}

The interaction of the cavity mode with an atomic beam gives rise
to a stochastic evolution, which is mainly due to the statistical knowledge of the number of atoms which have interacted with the cavity mode.
It is then appropriate to use a density-matrix formalism for the cavity-mode state. We denote by $\hrho$ the cavity-mode density matrix. The time evolution of the cavity field is characterized by: (i) the typical time scale which separates the arrival of two atoms, which is here given by the arrival rates $r$ and $r_e$ for the atoms prepared in state $|g\rangle$ and $|e\rangle$, respectively.  (ii) the interaction time $\tau$ between atom and resonator, which is determined by the atomic velocity and which follows a normal distribution $\mathcal P(\tau)$ (which we do not specify for the moment). Typically, $r\tau,r_e\tau\ll 1$ and one can study the field evolution on a coarse-grained time scale $\Delta t$, such that $\Delta t\gg \bar{\tau}$, with $\bar{\tau}$ the average interaction time. For $r\Delta t,r_e\Delta t\ll 1$  such that we can assume that there is at most one atom inside of the cavity, we can determine the density operator at the time $t+\Delta t$ given $\hat \rho(t)$, according to the formula
\begin{eqnarray}
\label{Master:0}
\hat \rho(t+\Delta t)&=& \hat \rho(t)(1-r\Delta t - r \mathcal R  \Delta t)\nonumber\\
&&+r\Delta t\int_0^{\infty} {\rm d}\phi\nonumber
p(\phi) \nonumber\\
&&\times \bigg[\cos(\phi \sqrt{\adag \hat
a})\hat \rho(t)\cos(\phi \sqrt{\adag \hat a})\nonumber\\
& &\left.+\frac{\sin(\phi\sqrt{\hat a\adag})}{\sqrt{\hat
a\adag}}\hat a\hat \rho(t)\adag\frac{\sin(\phi\sqrt{\hat
a\adag})}{\sqrt{\hat a\adag}}\right.\nonumber\\
& &\left.+\mathcal
R\cos(\phi \sqrt{
\hat a\adag})\hat \rho(t)\cos(\phi \sqrt{ \hat a\adag})\right.\nonumber\\
& &\left.+\mathcal R\adag\frac{\sin(\phi\sqrt{\hat
a\adag})}{\sqrt{\hat a\adag}}\hat \rho(t)\frac{\sin(\phi\sqrt{\hat
a\adag})}{\sqrt{\hat a\adag}}\hat
a\right]\nonumber\\
&\equiv&\hat \rho(t)+\Delta \hat \rho
\label{Delta:rho},
\end{eqnarray}
where we used $r_e=\mathcal Rr$ and we wrote the distribution $P(\tau)$ in terms of the distribution $p(\phi)$ of the Rabi angle $\phi=g\tau$. Operator $\Delta\hat \rho$ in Eq.~(\ref{Delta:rho}) is the differential change of the field state. The master equation is given by the differential equation $\partial \hat \rho_t/\partial t$ which is found from equation $\Delta\hat \rho/\Delta t$. This equation has the Lindblad form~\cite{Henkel}, as one can verify, but it has no trivial solution. Since the atoms are initially uncorrelated with the cavity mode, the inhomogeneous term of the Zwanzig-Nakajima master equation disappears \cite{Haake-book}. 
Moreover, the Markov approximation is valid in presence of a single cavity: the atoms exiting the resonators do not interact with it any longer, leading to no memory effects. 
Indeed, in \eqnref{Delta:rho} the density matrix at time $t+\Delta t$ depends only on $\hrho(t)$. 
We remark that, if the atoms exiting one resonator then interact with a second physical system, then correlations mediated by the atoms must be taken into account.

Below we discuss the master equation for the cavity mode in two specific limits: the weak-coupling regime, when the coupling of a single atom with the cavity mode is a small perturbation of the cavity state, i.e., $\phi\ll 1$, and the strong-coupling regime, when a single atom perturbs significantly the cavity state, and $\phi\ge 1$. From now on we denote by $\hat \rho_t$ the density matrix of the cavity field at time $t$ after the coarse-grained time averaging.

\subsection{The weak-coupling limit} \label{Sec:2:b}

The weak-coupling limit corresponds here to the regime in which the mean interaction time $\bar\tau$ fulfills the relation
$g{\bar\tau}{\sqrt n}\ll$ 1, where $n$ is any relevant photon number (such that the corresponding populations and coherences are not negligibly small) and the width of the distribution for $\tau$ is assumed to be small compared to the average value $\bar\tau$.
In this limit the field operators in Eq.~(\ref{Master:0}) can be expanded in powers of $\phi$, and
the dynamics of the cavity density matrix $\hrho_t$ is
governed by the master equation~\cite{master-eq} \begin{eqnarray}
\pdiff {\hat\rho_t}{t}&=&-{\rm i}\omega_c[\hat a^{\dagger}\hat a,\hat \rho]
          -\frac{\gamma}{2}\left(\adag \hat a\hat\rho_t -
          2\hat a\hat\rho_t\adag + \hat\rho_t \adag\hat a
          \right)\nonumber\\
         & & -\frac{\gamma_e}{2}\left(\hat a  \adag\hat\rho_t -
          2\adag\hat\rho_t\hat a + \hat\rho_t \hat a \adag
          \right),
          \label{eq:master}
\end{eqnarray} where $\gamma=rg^2\bar\tau^2$ and $\gamma_e=\mathcal R\gamma$ \cite{footnote}.
This equation
describes the incoherent energy exchange between the cavity field
and an external reservoir, with loss and pump rates $\gamma$ and $\gamma_e$, respectively.
If $\mathcal R<1$, the
resonator thermalizes with an effective reservoir at temperature
$$T=\frac{1}{\kappa_B}\frac{\hbar\omega_c}{|\ln \mathcal R|}.$$ The
steady state of the resonator is hence a thermal state, whose
temperature can be controlled by adjusting the parameter
$\mathcal R$, giving the average rate of atoms prepared in the
excited states over the ones prepared in the ground state.

The steady state can be a pure state, the vacuum state, by
preparing the atoms exclusively in the ground state $|g\rangle$
(which corresponds to setting $\mathcal R=0$ in
Eq.~(\ref{eq:master})): In this case the atoms absorb in average
energy from the cavity mode until it reaches the vacuum state
$|0\rangle$. This identifies a simple strategy which allows
one to prepare the cavity in the vacuum state as a result of the
interaction with a beam of atoms, of which one controls only the
initial internal state and the arrival rate. This procedure does not require
atomic detection nor control of the atomic velocity, but only the mean value and the variance of the velocity distribution, so to warrant
the weak-coupling regime. Following the line of reasoning presented in Ref.~\cite{PZnature}, this strategy could have also
been identified on the basis of the observation that the state
$|g,0\rangle$ is an eigenstate of Hamiltonian $\hat H$ such that
$\hat H_{\rm JC}|g,0\rangle=0$, and it is the unique dark state of
this dynamics. On this basis, one can construct a master equation
which has as steady state $|g,0\rangle$, and which has the form
given in Eq.~(\ref{eq:master}) for $\mathcal R=0$.

A useful benchmark for the quantum state preparation is given by the fidelity $\mathcal F (t)$ for preparing the system in the vacuum
state at time $t$ since the beginning of the experiment. The fidelity
$\mathcal F(t)=\langle 0|\hat \rho(t)|0\rangle$ corresponds
to the population of the vacuum state at time $t$. The solution
$\hat \rho(t)$ can be exactly evaluated using the damping
basis~\cite{Briegel}, and reads \begin{equation}
\hat \rho(t)=\sum_{n=0}^{\infty}{\rm e}^{-n\gamma
t}\hat \rho_n^{(0)}\alpha_n^{(0)}\end{equation} where $\hat \rho_n^{(0)}$
are the right eigenvector of the Liouvillean defined in
Eq.~(\ref{eq:master}) corresponding to the eigenvalue $n\gamma$, and
$\alpha_n^{(0)}$ is the $n$-th moment of the expansion in the
number operator, taken over the initial state of the cavity field.
Using the explicit form one
finds~\cite{master-eq} \begin{equation} \mathcal F(t)=1-\langle n\rangle_0{\rm
e}^{-\gamma t}+\frac{\langle n(n-1)\rangle_0}{2}{\rm e}^{-2\gamma
t}+\ldots \label{fidelity}\end{equation} where $\langle f(\adag \hat a)
\rangle_0={\rm Tr}\{f(\adag \hat a) \hat \rho(0)\}$. When the initial state
is thermal, with average photon number $\bar n=\langle \adag
\hat a\hat \rho(0)\rangle$, then expression~(\ref{fidelity}) takes the
compact form
\begin{equation}
\mathcal F_{\rm wc} (t)=\frac{1}{1+\bar n{\rm
e}^{-\gamma t}}=\frac{1}{1+\bar n{\rm e}^{-\bar \phi^2 r t}}
\label{fidelity:wc}
\end{equation}
and in this case one sees that the
time scale for reaching the ground state is determined by
the damping rate $\gamma$ and by the initial occupation number
$\bar n$. In other words, the achieved fidelity is given by initial occupation number $\bar n$, the average Rabi angle $\bar \phi$, and the average number of atoms $N=r t$ which have crossed the cavity.

\subsection{The strong-coupling limit}
\label{sec:strongcoupling}

The strong-coupling limit is characterized by $g \bar \tau  \ge 1$. In this regime, one can see from
Eqs.~(\ref{U:1}) and~(\ref{U:2}) that, for a given interaction time $\tau_0=\pi/g\sqrt{m_0}$ (and any integer multiple of $\tau_0$), with $m_0>0$
integer number, there is no net exchange of energy between cavity field and dipole when the initial state of the system is a coherent (or incoherent) superposition of the states $|g,m_0\rangle$ and $|e,m_0-1\rangle$. In other words, these states are fixed points of the resonator dynamics, in the absence of dissipation. They are however not unique: Indeed, the subspace of the fixed points of the evolution operator for a fixed interaction time $\tau_0$ has infinite dimension, being at least composed by all states $|g,m\rangle$, $|e,m-1\rangle$, with $m=\ell^2m_0$ and $\ell=1,2,\ldots$. These states have been denoted in the literature as trapping states~\cite{Meystre,WaltherTrapping}.

The theory of trapping states has been reported in~\cite{Meystre}. These properties have been used in milestone experiments~\cite{WaltherTrapping} in order to generate photon number states of the cavity field. The effect of noise on this dynamics have been theoretically analyzed in~\cite{Briegel:2,Andreas}. Using simple considerations we now discuss how the vacuum state of the electromagnetic field can be the unique asymptotic state of the dynamics, and determine the corresponding preparation fidelity.
For this purpose, we assume that all atoms are initially prepared in their ground state ($\mathcal R=0$) and that their velocity (interaction time) is distributed according to a function $p(\phi)$ with finite width. The time-evolution of the  diagonal elements of the field density matrix, $\langle n|\hat \rho(t)|n\rangle \equiv
c_n(t)$ (with $\sum_n c_n(t)=1$ as ${\rm Tr}\{\hat \rho\}=1$) is found from Eq.~(\ref{Master:0}), and is given by the set of coupled differential equations
\begin{eqnarray} \dot c_{n}(t) =-r B_n
c_{n}(t) + r B_{n+1} c_{n+1}(t)\label{eq:diffeq} \end{eqnarray} with
$B_0=0$ and \begin{eqnarray} B_n = \int_{0}^{\infty} {\rm
d}\phi\sin^2\left(\phi\sqrt {n}\right) p(\phi) \,. \label{Eq:Bn}\end{eqnarray}
Their solution reads \begin{eqnarray} c_0(t) &=&
c_0(0)+r B_1\int_{0}^t {\rm d}\tau c_1(\tau) \label{eq:sol_c_0}
\\
c_n(t) &=& {\rm e}^{-B_n r t} c_{n}(0) + r B_{n+1}
\int_0^t {\rm d}\tau {\rm e}^{-B_n r(t-\tau)}c_{n+1}(\tau),\nonumber
\end{eqnarray}
showing that the vacuum state is always a trivial stationary solution. Other stationary solutions can be found if there exist $n>0$ for which the coefficients $B_n=0$.

In order to study the behaviour of the coefficients $B_n$, let us assume that the velocities follow a Gaussian
distribution, such that $p(\phi)= \exp\left( -(\phi-\phi_0)^2/2 \sigma^2
\right)/\sqrt{2\pi\sigma^2},$ where $\phi_0$ is the center and
$\sigma$ the width. The coefficients $B_n$ then read
\begin{equation} B_n=\frac 1 2\left(1 - {\rm e}^{-2
n\sigma^2}\cos\left(2\phi_0\sqrt n \right)\right), \end{equation}
such that for $\sigma>0$ only $B_0$ vanishes. In this limit,
hence, the only stationary state is the ground state.
Nevertheless, for sufficiently small values of the width $\sigma$
there exist coefficients $B_n$, with $n>0$, whose value can be
very small, so that the occupation of the oscillator ground state
may converge very slowly toward unity as a function of time. A
limiting case is found when $\sigma\to 0$, such that $p(\phi) =
\delta(\phi-\phi_0)$.
This is the situation in which trapping states may exist.
For instance, if we choose $\phi_0=\pi/2$, then the coefficients
 $B_0$ and $B_{4n^2}$ (for $n=1,2,\ldots$) vanish,
indicating that the Fock states $|0\rangle$ and the states of the subset $\{|4n^2\rangle\}$ are
fixed points of the dynamics, i.e., trapping states. The initial state of the cavity determines in which of these trapping states the cavity will be found. In general the final state is a statistical mixture of these states.

The other limiting case is found for large $\sigma$, i.e. a broad distribution for the atom's velocity. For $\sigma \gtrsim \pi$, each of the coefficients samples approximately equally over the period of the sinusoidal function in \eqnref{Eq:Bn}, avoiding trapping states, and we can therefore set $B_n=\frac 1 2$. For the fidelity in this approximation we find
\begin{eqnarray}
\mathcal F_{\rm sc}(t)=1-e^{-r_{\rm
at}t/2}\sum_{n^=0}^{\infty}c_{n}(0)\sum_{m=0}^{n-1}\frac{(r t/2)^m}{m!} \label{c_00 for equal proba2},
\end{eqnarray}
where we used the relation $\int_{0}^{t}\tau^n e^{-D\tau}{\rm d}\tau =\frac{n!}{D^{n+1}}\left[1-e^{-Dt}\sum_{m=0}^{n}\frac{(Dt)^m}{m!}\right]$. For an initial thermal state with $c_n(0)=(1-\mu^2)\mu^{2n}$, where $\mu$ is given by the initial average number of thermal photons via $\left<\adag \hat a \right>=\frac {\mu^2}{1-\mu^2}$, this simplifies to
\begin{equation}
\mathcal F_{\rm sc}(t) = 1 -\mu^2 e^{-r t(1-\mu^2)/2},
\label{eq:Fsc}
\end{equation}
and one sees that in this limit the fidelity of the ground state is determined only by the initial occupation number $\bar n$ (parametrized by $\mu$) and by the number of atoms $r t$ which have crossed the cavity.

\section{Generating EPR Entangled Radiation with an atomic
reservoir} \label{Sec:3}

In the following we discuss in detail and extend a proposal for
generating EPR-entangled states of two cavity modes by resonant
interaction with a beam of atoms, which was first presented in Ref.~\cite{Pielawa07}. The setup is sketched in Fig.~\ref{Fig:exp-setup}(a), and is composed by a resonator,
with two modes in which a single atom can emit photons. The atoms are part of a beam crossing the resonator,
and are driven by a classical microwave field while they interact with the cavity modes. The interaction is shaped in such a way
that the atoms emit correlated photons into the cavity modes. The underlying
mechanism is four-wave mixing, where emission into the cavity
modes is enhanced by resonant coupling with the Rabi sidebands
induced by a classical field that saturates the atomic
transition~\cite{Agarwal}, with the creation of EPR-correlations
being enforced by the initial quantum state of the injected atoms.
Contrary to typical setups based on optical parametric amplifiers,
the atoms pump the resonator through resonant single-photon
processes.

Before discussing details of the mechanism, let us make some
general remarks on the main ideas regarding the generation of EPR states through reservoir engineering.
We first label by $1$ and $2$ the two cavity
modes, such that $\omega_j$ is the  frequency and $\hat a_j$, $\hat a_j^{\dagger}$ are the annihilation and creation operators
of an energy quantum $\hbar\omega_j$ of the corresponding mode
($j=1,2$). Our goal is to generate a dynamics described by a
master equation whose steady state is a two-mode squeezed state,
$\hat\rho_{\rm St}=|\psi\rangle\langle\psi|$, with
\begin{equation} |\psi\rangle=\hat S^\dag(\xi)\ket{0,0}.
\end{equation}
Here, $|0,0\rangle$ is the vacuum state of both cavity modes and
\begin{equation} \hat S(\xi) = \exp\left(\xi ^* \hat a_1 \hat a_2
- \xi \adag_1\adag_2 \right) \label{eq:squeezing} \end{equation}
is the two-mode squeezing operator, with $\xi$ a complex parameter.
It can be verified that state $|\psi\rangle$ is the vacuum state of
harmonic oscillators, whose annihilation operators $\hat b_1$ and $\hat b_2$
are related to the cavity modes operators by the relation
\begin{subequations}%
\begin{eqnarray}%
&&\hat b_1= \hat S^\dag(\xi) \hat a_1 \hat S(\xi),
\label{eq:b1operator}
\\&&
\hat b_2= \hat S^\dag(\xi) \hat a_2 \hat S(\xi),
\end{eqnarray}%
\label{eq:boperators}%
\end{subequations}%
and analogously for the creation operators. 
We refer to modes $\hat b_1$ and $\hat b_2$ as Bogoliubov modes, in analogy with the Bogoliubov transformation used in solid-state physics \cite{Bruus_Flensberg}. 
State $\hat\rho_{\rm St}$
can be hence the dark state of an interaction Hamiltonian $\hat H_1$,
which has, say, the form of the Jaynes-Cummings Hamiltonian
$\hat H_{\rm JC}$ in Eq.~(\ref{Eq:JC}) but with the operator $\hat a$
($\adag$) replaced by the operator $\hat b_1$ ($\hat
b_1^{\dagger}$). We note that it is also the dark state of the
interaction Hamiltonian $\hat H_2$, which has the same form as $\hat H_1$
but now with $\hat b_2$ and $\hat b_2^{\dagger}$ in place of $\hat b_1$ and $\hat b_1^{\dagger}$. In particular,
$\hat \rho_{\rm St}$ is simultaneously dark state of both interactions, and it is unique.
Hence, a dynamics can be constructed, which has as unique steady state $\hat \rho_{\rm St}$, by implementing sequentially
two interactions which effectively damp oscillators $\hat b_1$ and $\hat b_2$, respectively. In the following we will show how to engineer such dynamics.

\subsection{Engineering the coupling to the reservoir}

\begin{figure}[t] \includegraphics[width=0.4\textwidth]{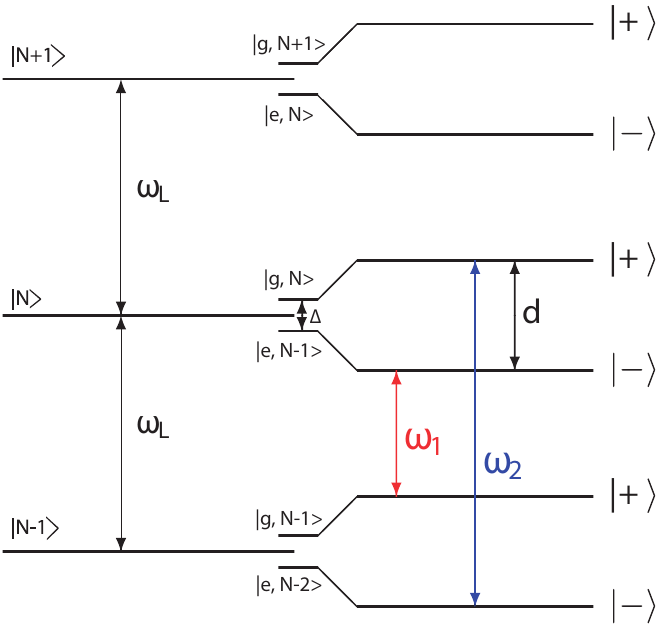}
\centering
\caption{  Left ladder: energy of photon states $|N\rangle$ (with $N\gg 1$) of the semiclassical field at frequency $\omega_L$. Middle ladder: corresponding energies of the doublets of states $\{|g,N\rangle,|e,N-1\rangle\}$, where $\Delta=\omega_L-\omega_0$ is the corresponding splitting in energy. Right ladder: energy of the semiclassical dressed states $\ket \pm$, Eq.~(\ref{dressed:states}), with energy splitting $d$, Eq.~(\ref{Eq:d}). A transition $\ket + \to \ket -$ is accompanied by absorption (emission) of a photon of frequency $\omega_1$ ($\omega_2$) from (into) the corresponding cavity mode.} \label{Fig:ladder}
\end{figure}

The Hamiltonian of driving field, atom, and cavity modes, in the reference frame rotating at the frequency $\omega_L$ of the classical field, has the form
\begin{eqnarray} \hat H^{RF}=\hat {\cal
H}_0- \sum_{j=1,2}\hbar\delta_j\adag_j\hat a_j +\sum_{j=1,2} \hbar
g_j \left( \sigmdag \hat a_j + \sigm \adag_j \right),
\label{eqn:Hamiltonian-laser-frame}
\end{eqnarray}
where $g_j$ are
the coupling constants between the two-level atom and each cavity
mode, detunings $\delta_j=\omega_L-\omega_j$, $\Delta= \omega_L - \omega_0$, and \begin{equation} \hat
{\cal H}_0=- \hbar \Delta \sigmdag\sigm + \hbar\Omega
\left(\sigmdag + \sigm\right) \end{equation} describes the
coupling between dipole and classical field, with strength $\Omega$, see Fig.~\fref{Fig:ladder}.

Let the coupling to the classical field be much stronger than the coupling to the cavity modes,
$|\Omega| \gg |g_\lambda|$, it is then convenient to express Hamiltonian~(\ref{eqn:Hamiltonian-laser-frame}) in the basis of eigenstates $\ket{\pm}$ of $\hat {\cal H}_0$, with
\begin{equation} \hat {\cal H}_0\ket{\pm}=-\hbar(\Delta\mp
d)/2\ket{\pm}, \end{equation} and \begin{equation} \label{Eq:d} d =
\sqrt{\Delta^2 + 4\Omega^2}\,.\end{equation} The states $|\pm\rangle$
are the semiclassical dressed states, and read
\begin{subequations}
\begin{eqnarray}
\ket{+}&=&\sin\theta\ket g + \cos\theta\ket e,\\
 \ket {-} &=&\cos\theta\ket g - \sin\theta\ket e,
\end{eqnarray}
\label{dressed:states}
\end{subequations}
with
\begin{equation}
\label{theta}
\tan\theta = \frac{2|\Omega|}{d-\Delta}\,.
\end{equation}
The corresponding energy levels are shown in Fig.~\fref{Fig:ladder}.
We introduce the
raising and lowering operators for the dressed states basis,
\begin{equation} \piplus = \ket +\bra -,\qquad \piminus = \ket
-\bra +, \end{equation} with $\piz = \ket +\bra + - \ket -\bra -$.
Using the semiclassical dressed-state basis and operator notation
we rewrite Eq.~(\ref{eqn:Hamiltonian-laser-frame}) as $\hat H^{RF}
= \hat H_0 + \Hint$, with $${\hat H_0 = \hbar d\piz/2
-\hbar\sum_{\lambda}\delta_{\lambda} \adag_{\lambda}\hat
a_{\lambda}},$$ and
\begin{eqnarray} \label{eqn:Hint-total} &&\Hint
= \sum_{\lambda}\hbar g_\lambda \left[ \piz\left(\hat a_\lambda
+\adag_\lambda\right)\cos\theta\sin\theta \right.\\&&
\left.+\left(\piplus\hat a_\lambda
+\adag_\lambda\piminus\right)\cos^2\theta - \left(\piminus\hat
a_\lambda+\piplus\adag_\lambda\right)\sin^2\theta\right].
\nonumber \end{eqnarray}
If $|g_\lambda|\ll d$ we can choose which
processes are resonant, and thus relevant for the dynamics, by
changing the values of $\delta_1, \delta_2$, and $d$. One is thus
able to generate a diversity of dynamical processes, some of which are discussed in Refs.~\cite{Solano,Pielawa07}.\\

Let us set $\delta_1 = d$ and $\delta_2 = -d$, as shown in Fig.~\fref{Fig:ladder}.
If $|g_{\lambda}|\ll d$ we obtain from~\eqnref{eqn:Hint-total} the effective Hamiltonian $\hat H^{RF}\approx \hat
H_{\rm eff}=\hat H_0 + \Hint$, with
\begin{equation} \hat H_{\rm int}=
\hbar g\left(\adag_2 \cos^2\theta-\hat a_1
\sin^2\theta\right)\piminus+\hc\,, \label{eqn: Hinteff}
\end{equation}
and $\hat H_0 = \hbar d(\piz/2 - \adag_1\hat
a_1+\adag_2\hat a_2)$. We have assumed $g:=g_1=g_2$. The
processes described by Eq.~(\ref{eqn: Hinteff}) are indicated by
the arrows in Fig.~\fref{Fig:ladder}. 
In the basis of the $b$ operators, we have
$\hat H_0=\hbar d
\left(\piz/2-\hat b^\dag_1\hat b_1+\hat b^\dag_2\hat b_2\right)$
and
\begin{subequations}
\begin{eqnarray}
&&\Hint^1=-\hbar\OmBog \left( \hat b_1 \piminus + \hat b^\dag_1\piplus\right) ,
{\rm ~~if~~~} \Delta>0\,
\label{eqn: Hint in bog basis+}\\
&&\Hint^2=\hbar\OmBog\left( \hat b^\dag_2 \piminus + \hat b_2\piplus \right) ,
{\rm ~~if~~~}\Delta<0\,. \label{eqn: Hint in bog basis-}
\end{eqnarray}%
\label{eq:SqJCint}
\end{subequations}%
Here, $$\OmBog = g \sqrt{(1-\mu)/(1+\mu)}$$ with
$r_\mu=\arctanh\mu$, while the value of $\mu$ is determined by the
classical field parameters, \begin{eqnarray} \label{mu}
&&\mu=\tan^2\theta~~ {\rm if}~~\abs{\tan\theta} <
1\\&&\mu=(\tan\theta)^{-2}~~ {\rm if}~~\abs{\tan\theta} >
1.\nonumber\end{eqnarray}
We now can see that the interaction of the cavity modes with a beam of atoms, each initially prepared in the state  $\ket +$ ($\ket -$) and undergoing the dynamics governed by Hamiltonian $\hat H_1=\hat H_0+\Hint^1$ ($\hat H_2=\hat H_0+\Hint^2$), will give rise to an effective dynamics, whose steady state is the pure state $\ket\psi$ of the cavity modes. The Hamiltonian governing the dynamics will be $\hat H_1$ or $\hat H_2$ depending on the sign of the detuning $\Delta$, which can be controlled by appropriately shifting the atomic transition frequency.

\subsection{Effective dynamics: reaching the EPR state}

\begin{figure}[t] \includegraphics[width=0.45\textwidth]{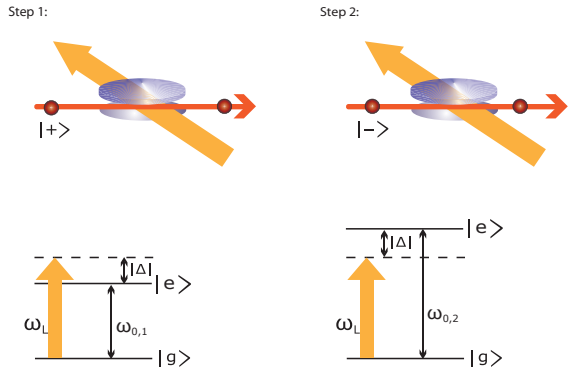}
\caption{  Schematic representation of the interaction processes needed, in order to prepare the cavity modes in a EPR state. In step 1, the atoms are prepared in the state $\ket{+}$ before crossing the resonator. Inside the resonator they are driven by a classical field, and the atomic frequency is shifted with respect to $\omega_L$ according to the energy level scheme displayed at the bottom of the figure. In step 2, the initial state is $\ket{-}$ and the atomic transition frequency is shifted, such that the detuning with the field has opposite sign.}
\label{Fig:two-step}
\end{figure}

We discuss now how the cavity modes can be prepared in the two-mode squeezed state asymptotically. This is achieved by an effective ``{\it dissipation}" process in the {\it b}-basis, implemented in a two-step procedure sketched in Fig.~\ref{Fig:two-step}. The first step consists in letting the atomic beam interact with the resonator with each atom prepared in the state $|+\rangle$ and the detuning of the classical field set to the value $\Delta=\Delta_0>0$.
Inside the cavity each atom undergoes the dynamics of Eq.~(\ref{eqn: Hint in bog basis+}), such that at the end of the interaction, on average, excitations have been removed from  mode $\hat b_1$. In the second step, the atoms are prepared in state $\ket -$ and the detuning of the classical field is set to the value $\Delta= -\Delta_0$. Inside the resonator each atom undergoes the dynamics given by Eq.~(\ref{eqn: Hint in bog basis-}), such that at the end of the interaction, on average, excitations have been removed from  mode $\hat b_2$.

In the weak-coupling regime, the equation of motion of the cavity field modes is given by the master equation for the density matrix $\hrho_t$, which during each step $j$ ($j=1,2$) reads
\begin{equation}
\left.\pdiff{\hrho_t}{t}\right|_\textrm{\textrm{step j}} = -
\frac{\gamma}{2} \left(\bdag_j\hat b_j \hrho_t - 2\hat
b_j\hrho_t\bdag_j+\hrho_t\bdag_j\hat b_j \right),
\label{master:eq} \end{equation} where
\begin{equation}\label{gamma}\gamma=\rate\OmBog^2\tau^2\,,
\end{equation}
$\tau$ is the interaction time, and $\rate$ is the atomic arrival
rate. Correspondingly, during each step the average number of Bogoliubov excitations is exponentially damped according to the equation $\ave{\bdag_j\hat b_j}_t = \ave{\bdag_j\hat b_j}_{0} \exp(-\gamma t)$, and vanishes at
times $t \gg 1/\gamma$ (see also Sec. \ref{Sec:2:b}).
In terms of the original field modes, this procedure implies that the atoms pump in phase only the two-mode squeezed state. Asymptotically, after the implementation of the two steps for a sufficiently long time (such that the cavity field is still stable over this time) the field state approaches the state
\begin{equation} \hrho_\infty =
\ket{0,0}_b\bra{0,0}=\Stdag(r_\mu)\ket{0,0}_a\bra{0, 0}\St(r_\mu),
\end{equation} which is a two-mode squeezed state, and whose degree of
squeezing $r_\mu$ is solely determined by the ratio $|\Delta/\Omega|$. This state is reached independently of the
initial state of the cavity modes, provided that each step is implemented for a sufficiently long time $T$.

An analogous dynamics can be implemented in the strong-coupling regime, i.e., when $\tau\Omega_{\rm b} \gtrsim 1$. Here, one aims at creating a trapping state in the Bogoliubov basis for each Bogoliubov oscillator in a two-step procedure, where the initial state of the atomic beam and the atomic parameter are changed as in the procedure outlined for the weak-coupling case. In this case, the creation of trapping states in the Bogoliubov basis, with numbers $n_1$ and $n_2$ for each Bogoliubov mode, corresponds to entangled states of the cavity modes of the form $\hat S(\xi)|n_1,n_2\rangle$. Choosing a broad velocity distribution, one finds that the steady state is a vacuum two-mode squeezed state, $\hat S(\xi)|0,0\rangle$.

\subsection{Experimental parameters}

\indent
The proposal we discussed so far is based on a stochastic dynamics, where the average action of each atom leads the cavity modes to a stationary, EPR-entangled state. The  scheme does not require atomic detection, nor control of the number of atoms, nor of the interaction times (atomic velocities).  A possible experimental setup of this scheme is sketched in Fig.~\ref{Fig:exp-setup}(a). Prior to the interaction region, the atoms are prepared in a coherent superposition of two Rydberg states $\ket g$ and $\ket e$ connected by a dipole transition. Inside the resonator a classical field saturates the dipole transition, thereby pumping on resonance the two nondegenerate modes of the resonator, as shown in Fig.~\ref{Fig:ladder}.

While the interaction time between cavity and each atom needs not to be controlled, on the other hand the dynamics between atom and cavity is here assumed to be Hamiltonian, and characterized by a two-level transition which can be tuned on resonance with the cavity modes by means of an external field. Let us now discuss these assumptions individually. Selecting a two-level transition, here denoted by the electronic states  $|g\rangle\to|e\rangle$ imposes constrains on the field polarizations, such that they all couple with
the dipole transition, while coupling to other states is avoided. In an open-cavity geometry~\cite{Haroche-Colloquium}, this can be achieved by means
of an electric potential between the two mirrors, which removes through Stark shifting the degeneracy of circular Rydberg states,
and using circular polarizations for both the cavity modes and the pump field. The two-step procedure needs a change in the transition frequency
of the two-level atom, which can be achieved by an external static field.

The assumption of Hamiltonian dynamics between individual atoms and resonator relies on the fact that the atom must not decay during the interaction with the cavity modes, and dissipation of the cavity field should be negligible during the experiment. Experiments with microwave resonators~\cite{Haroche-Colloquium,CavityQED-Walther} are characterized by interaction times of the order of tens of $\mu$s, which warrant negligible spontaneous decay, typically of the order of tens of ms. The requirement that the cavity does not decay over the duration of the experiment $T_{\rm tot}=2T$ is instead more delicate, as it requires that the resonator is stable over the total interaction time with the atomic beam. Moreover, the time $T$ must be sufficiently smaller than the coherence time of the driving fields, so that the amplitude $\Omega$ remains constant.

\begin{figure}[t] \includegraphics[width=0.45\textwidth]{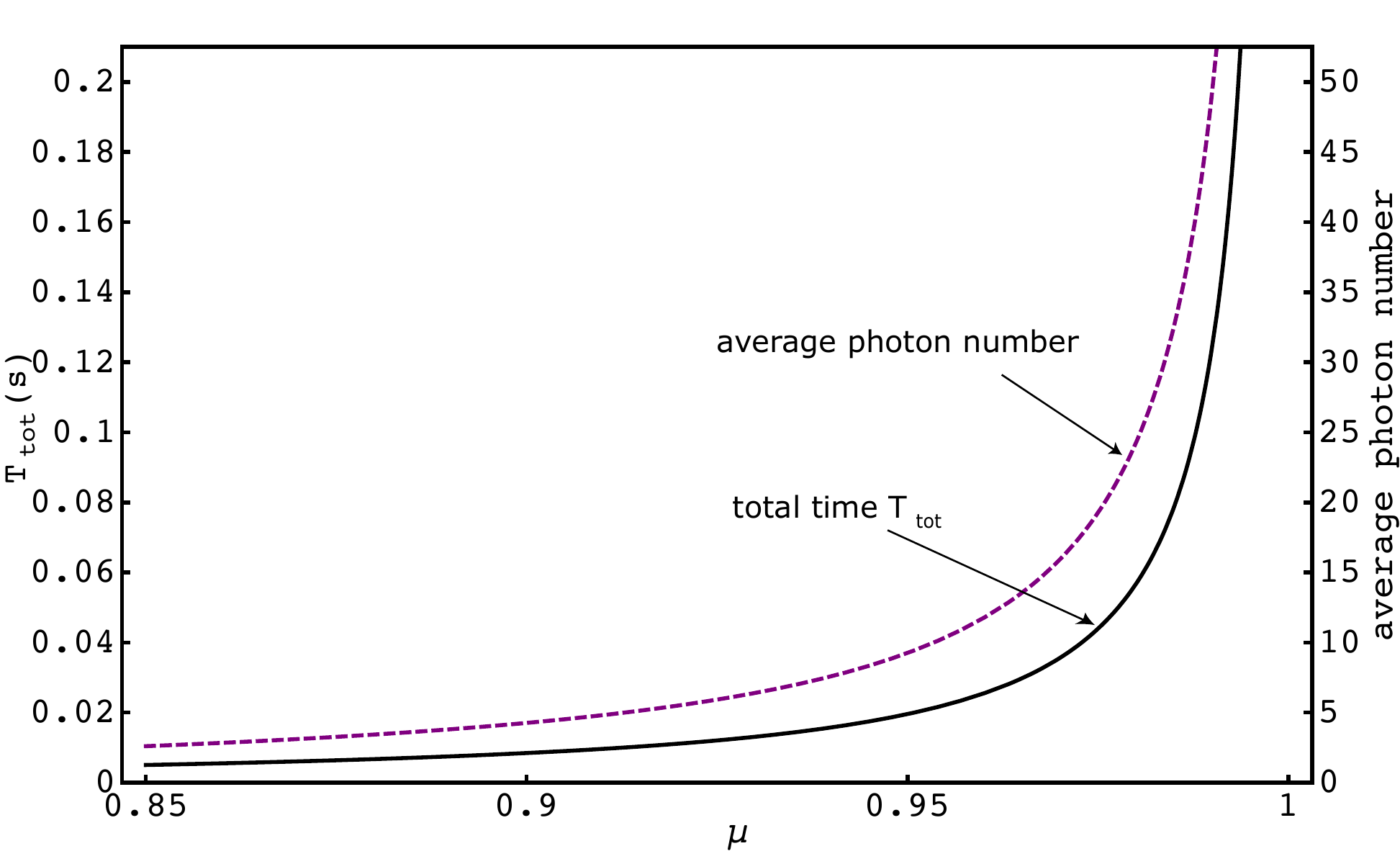}
\caption{  
(i) Solid line: Total experimetal time $T_{\rm tot}=2T$ (in seconds) and (ii) Dashed line: average photon number per mode at the end of the protocol as a function of $\mu$ (which is controlled by the intensity of the classical field and the detuning according to Eq. (29) and (24)).  The parameters we used are effective coupling $g =$~125~kHz, interaction time $\bar\tau=12.5{\rm \mu s}$, atomic arrival rate 11200 atoms/sec.
}
\label{Fig:photon_number_exp_time}
\end{figure}

Being the dynamics stochastic, the desired EPR-state is reached asymptotically. For this purpose, we estimate the time needed for reaching a given fidelity in the preparation of the desired state. We first focus on the weak-coupling limit. Using Eq.~(\ref{fidelity:wc}) we find that the fidelity of the protocol has the form
\begin{equation} \label{Fid:EPR:wc} \mathcal F_{wc}(T_{\rm tot})=\prod_{j=1,2}
\frac 1 {1+\ave{b^\dag_j \hat b_j}_0 e^{-\gamma T}} \end{equation}
where we have assumed that each step is performed on the time interval $T$, such that the total duration of the protocol $T_{\rm tot}=2T$. Clearly, the initial state affects the time scale required for reaching the desired fidelity. In particular, when the cavity modes are initially in the vacuum state one has $\ave{\bdag_j\hat b_j}_0=\frac {\mu^2} {1-\mu^2}=:\bar n_0$. When they are in a thermal state with $n_{th}$ thermal photons in each mode, instead, thus $\ave{\bdag_j\hat b_j}_0=n^{th}_0=\bar n_0+n_{th}+2 n_{th}\bar n_0$.
Figure~\ref{Fig:photon_number_exp_time} displays
the estimated total experimental times and corresponding average
number of photons per mode at steady state as a function of $\mu$,
where $\bar{n}_0=\mu^2/(1-\mu^2)$ when the cavity modes are in the
vacuum state at $t=0$. For the degree of squeezing
$r_\mu\approx2.1$ ($\mu=0.97$), leading to an average number of 16
photons per mode at steady state, and $\bar{n}_\infty=0.01$,
corresponding to a fidelity ${\cal F}\approx 0.98$, then one has $T_{\rm tot}\sim 36$ ms in case of an initially empty cavity ($T_{\rm tot}\sim 43$ 
ms for
0.7 thermal photons). Resonators stable over this time are
available in present experiments~\cite{ultra}. From these results we also see that fluctuations in the
coupling with the driving field, $\delta \Omega$, are negligible with current microwave sources.

We now consider the implementation of the protocol in the strong-coupling limit. In this case the fidelity for preparing the cavity modes in the two-mode squeezed state takes the form
\begin{equation}
\mathcal F_{sc}(T_{\rm tot})=\left(1-\mu^2 e^{-r_\at T (1-\mu^2)/2}\right)^2,
\label{Fid:EPR:sc}
\end{equation}
where we have used \eqnref{eq:Fsc}, assuming a broad distribution of atomic velocity and an initially empty cavity. 

\begin{figure}[t]
\includegraphics[width=0.4\textwidth]{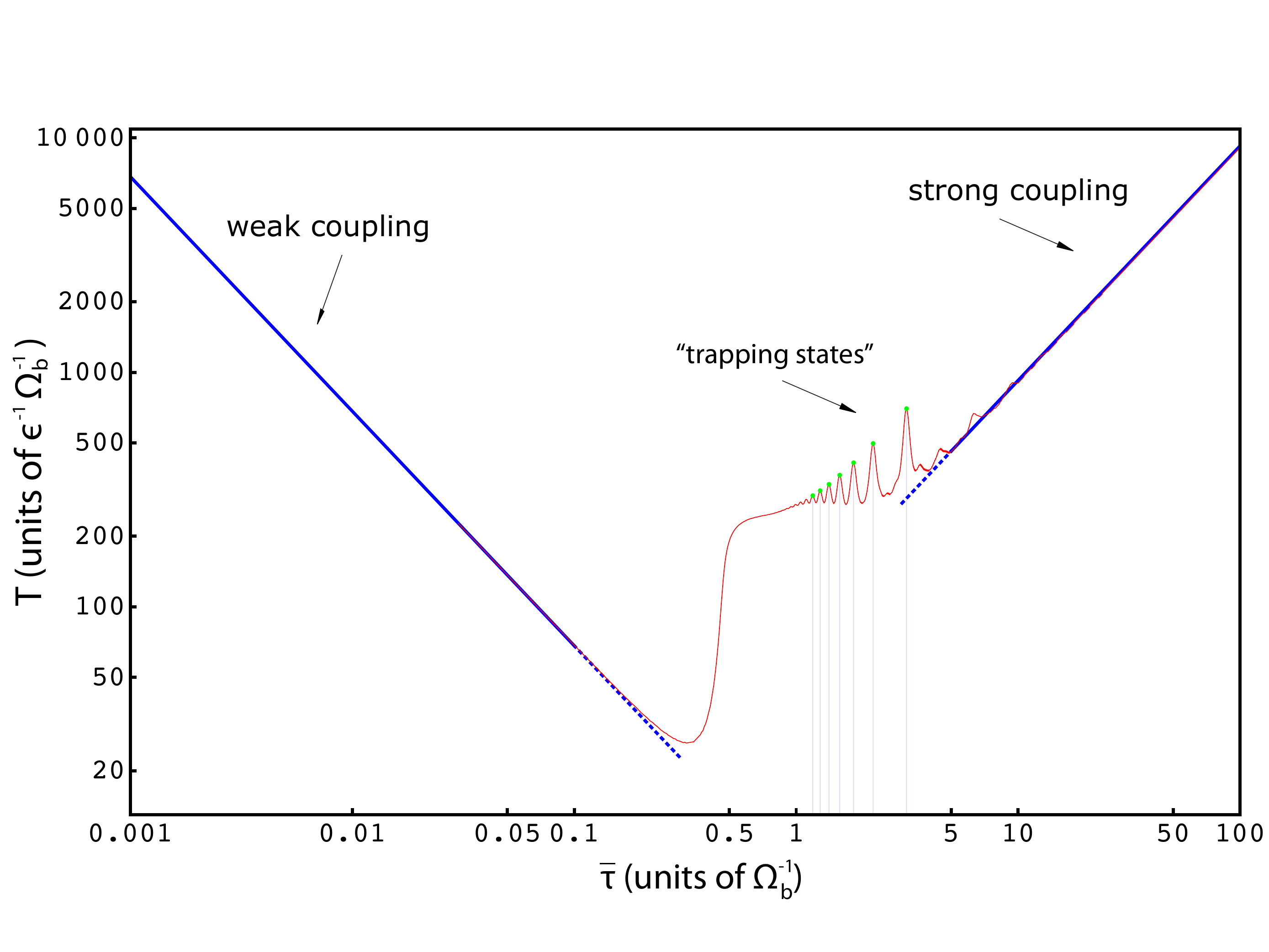}
\centering
\caption{  
Logarithmic plot of time T required in each step to reach the EPR state as a function of average interaction time $\bar \tau$ (in units of $\Omega_{\rm b}^{-1}$). The EPR state here chosen is characterized by $\mu=.95$ ($\bar n_0=9.3$), and we require that the corresponding fidelity at $T_{\rm tot}=2T$ is ${\cal F}=0.99$. The blue lines represent analytical expressions for fidelities $\mathcal F_{\rm wc}$, Eq.~(\ref{fidelity:wc}), and $\mathcal F_{\rm sc}$, Eq.~(\ref{eq:Fsc}). The red line has been obtained from numerical solution of \eqnref{eq:diffeq}, where coefficients $B_n$ were calculated taking $\sigma = 0.05 \phi_0$ in the Gaussian distribution function of Rabi angles $\phi_0=\bar\tau \Omega_b$. The most efficient regime, in terms of fastest protocols for a given fidelity, lies between weak and strong coupling. The plot shows slower convergence at Rabi angles corresponding to $\phi_0 = \pi / \sqrt{j}$ ($j$ positive integer) due to trapping states.
}
\label{Fig:swcoupling}
\end{figure}

In order to compare the efficiency of the two procedures, we now evaluate the time required to reach a desired fidelity in the weak and in the strong-coupling limit. We first observe that the system accesses the strong-coupling regime by increasing the Rabi angle $\phi$. In the case here considered, $\phi$ can be changed by changing the interaction time while keeping $g$, and thus $\Omega_{\rm b}$ fixed. Note that increasing the interaction time implies that the atomic arrival rate $r_\at$ needs to be adjusted to avoid simultaneous presence of more than one atom in the cavity. Let us therefore take the average number of atoms in the interval of time $\tau$ so that $\epsilon=r_\at\bar\tau \ll 1$. In the weak-coupling limit the fidelity in Eq.~(\ref{Fid:EPR:wc}), shows that it is favorable to increase $\bar\tau$ even at the expense of a slower rate $r_\at$ (see
Eq.~(\ref{gamma})). In the strong-coupling limit, from Eq.~(\ref{Fid:EPR:sc}) we find that increasing $\bar \tau$ further gives slower convergence to the desired fidelity, hence slower protocols. Figure \fref{Fig:swcoupling} displays the time required for reaching a desired fidelity $\mathcal F=0.99$ as a function of $\bar\tau$. One finds
an optimal $\bar \tau$, for which the protocol is fastest, in between the two regimes.

\section{Entangling two distant cavities with an atomic reservoir}
\label{Sec:4}

Thus far we have considered two modes of the same microwave cavity. However, as EPR entangled radiation is a possible resource for quantum communication with continuous variables, e.g. for quantum teleportation~\cite{Braustein04}, it would be desirable to create two entangled modes belonging to two different, spatially-separated cavities. In this section we extend the concepts discussed in Section~\ref{Sec:3} and present a scheme for entangling the
modes of two spatially separated resonators using an atomic beam.

We assume two open resonators, which are crossed by a beam of atoms propagating along the $z$ axis and whose relevant modes are at frequency $\omega_1$ and $\omega_2$, with annihilation and creation operators $a_1,a_1^{\dagger}$ and $a_2,a_2^{\dagger}$, respectively. The resonators spatial mode functions along the $z$ axis are Gaussians centered at $z_1$ and $z_2$, respectively, such that the distance $|z_1-z_2|=D$ is much larger than the modes waist $w$, namely, $D\gg w$ and the fields mode functions have no spatial overlap. The atomic transition is quasi-resonant with a mode at frequency $\omega_1$ in the first resonator, and at frequency $\omega_2$ with the second resonator. In addition, the atoms are driven by a maser which propagates almost parallel to the $z$-axis and which has wave vector $k$, frequency $\omega_L$ and intensity $\Omega$ (we neglect any spatial gradient and assume that the maser intensity is uniform along $z$).

The Hamiltonian describing the coherent interaction of one individual atom of the beam with both cavities reads \begin{eqnarray}
\hat
H(t)&=&\hbar\omega_0\hat\sigma^\dagger\hat\sigma\\
&&+\hbar\Omega\left({\rm e}^{-{\rm i}(\omega_Lt-kz(t))}\hat\sigma^\dagger+{\rm e}^{{\rm i}(\omega_Lt-kz(t))}\hat\sigma\right)\nonumber\\
&&+\sum_\lambda\left[\hbar\omega_\lambda\hat a^\dagger_\lambda
\hat a_\lambda+\hbar g_\lambda f_{\lambda}(z(t))(\hat
a_\lambda\hat\sigma^\dagger+\hat
a^\dagger_\lambda\hat\sigma)\right]\,, \nonumber\end{eqnarray}
where $g_\lambda$ is the strength of the coupling between cavity mode and dipolar transition, while the spatial mode function takes the form $$f_\lambda(z)=\exp(-(z-z_\lambda)^2/2w^2)/\sqrt{2\pi w^2},$$ with $w$ the mode waist. The atomic center of mass is located at the time-changing position $z(t)=z(0)+vt$, where $v$ is the velocity of the atom. The assumption of classical center-of-mass motion is justified by the parameters of the typical experimental situation, where the atoms exit an oven and the velocity selection brings to distributions still within the classical regime. In addition, the assumption of uniform motion is not necessary, but convenient for the theoretical treatment. Indeed, as we will show, the atomic velocity can change without affecting the efficiency of the protocol.

For the following treatment we assume that the mode waist is much smaller than the wavelength, as it is often the case, so that the maser field has a well defined phase over the interaction region. In particular, we denote by $\psi_1=kz_1$ and $\psi_2=kz_2$ the phases at each resonator. The atom is continuously driven by the maser field, and the phase of the dressed states is here assumed to follow adiabatically the phase of the field as the atom moves. For simplicity, in what follows we assume the phases $\psi_1=\psi_2=0$. Within these approximations, we write the effective dynamics of the individual atom interacting sequentially with the cavity modes in terms of semiclassical dressed states of the maser field, given in Eqs.~(\ref{dressed:states}). Setting the detuning between maser and cavity modes $\omega_L-\omega_1=d$ and $\omega_L-\omega_2=-d$, with $d$ the frequency splitting between the dressed states given in Eq.~(\ref{Eq:d}), we obtain the corresponding effective Hamiltonian, which reads $\hat H_{{\rm eff},j} = \hat H_0+\hat H_j$ with
\begin{subequations} \begin{eqnarray}
\hat H_0&=&\hbar d \left[\frac{\piz}{2}-\adag_1\hat a_1 + \adag_2\hat a_2\right]\label{H_zero},\\
\hat H_1&=&-\hbar g_1\sin^2 \theta f_1(z)\left[\adag_1\piplus+\hat a_1\piminus\right]\label{H_1},\\
\hat H_2&=&\hbar g_2\cos^2\theta f_2(z)\left[\adag_2\piminus+\hat
a_2\piplus\right]\label{H_2},
\end{eqnarray}%
\label{Hamiltonian}%
\end{subequations}%
where the angle $\theta$ is given in Eq.~(\ref{theta}). The space-dependence of the cavity spatial mode functions can be substituted by the mean value of the function over the interaction region, which is constant over an interval of time, such that the total pulse area is preserved~\cite{Englert:Lectures}.

Hamiltonian \eqnref{Hamiltonian} has the same form as \eqnref{eqn: Hinteff} in Sec.~\ref{Sec:3}, with the only difference that the atom interacts with each mode in well-separated time intervals, due to the distant location of the cavities. We will now show that the combined state of the two distant cavity modes can be pulled into a two-mode squeezed state as a result of the interaction with an atomic beam, of which the number of atoms and the individual interaction times are known only statistically.

\begin{figure}[tb] \centering
\includegraphics[width=0.45\textwidth]{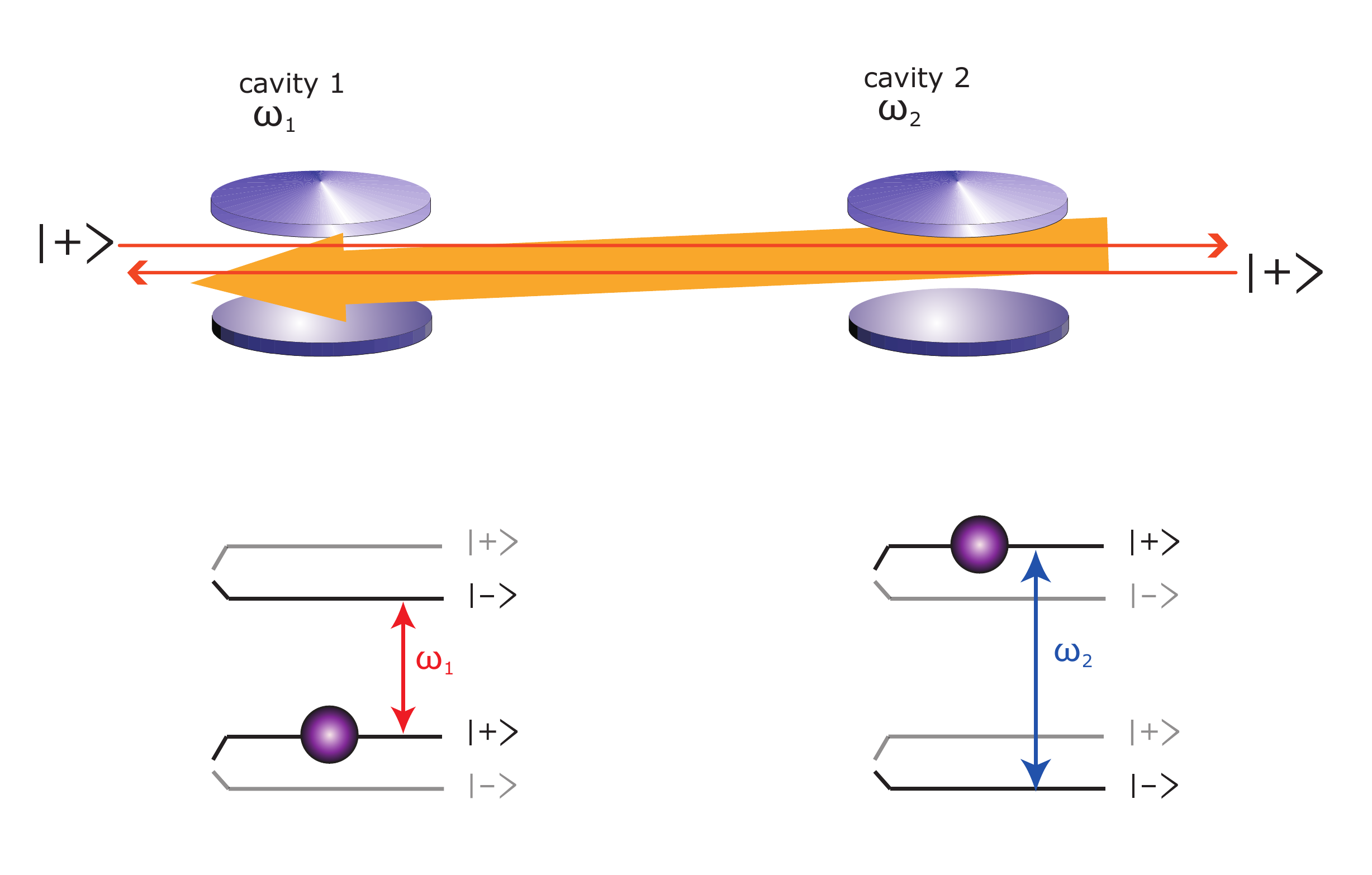}
\caption{  Top: sketch of the setup for entangling the modes of two distant resonators. Two atomic beams propagate in both directions and cross the resonators, interacting sequentially with each of them. Here one of the two required steps of the procedure is shown, where the atoms are prepared in the state $\ket{+}$. Bottom: energy levels which are coupled inside each resonator. The ``ball" represents the atomic occupation. 
An atom traveling from right to left first emits a photon $\omega_2$ into the second resonator, then emits a photon $\omega_1$ into the first one, only if it has previously emitted into resonator 2, otherwise it may absorb a photon $\omega_1$. An atoms traveling from left to right absorbs/emits photons in the reversed sequence. 
} \label{fig:two_beams} \end{figure}

The line of reasoning extends the protocol presented in Sec.~\ref{Sec:3}. In this case, however, one must consider that the interaction of the atom with each mode is sequential. For this purpose we introduce the evolution operator  $\hat U_j(\tau)$ for the dynamics of the atom interacting with the resonator $j$ over the interval of time $\tau$ (here taken in interaction picture with respect to Hamiltonian $\hat H_0$ in Eq.~(\ref{H_zero})) and we evaluate the total density matrix $\hat \rho_t^{\rightarrow}$ of atom and resonator after the atom has interacted with both cavities in interaction picture, assuming that the atom interacts sequentially first with mode 1 and then with mode 2. Denoting by $\Delta t$ the interval of time in which this occurs, the density matrix after the interaction reads
\begin{equation}
\hrho^{\rightarrow}(t+\Delta t)=\hat U_2(\tau)\hat U_1(\tau)\hrho(t)\hat U_1^{\dagger}(\tau)\hat U_2^{\dagger}(\tau).
\end{equation}
In order to obtain the two-mode squeezing correlations, we need also processes in which the temporal sequence of photon absorption and emission is reversed, namely, processes of the sort
\begin{equation}
\hrho^{\leftarrow}(t+\Delta t)=\hat U_1(\tau)\hat U_2(\tau)\hrho(t)\hat U_2^{\dagger}(\tau)\hat U_1^{\dagger}(\tau)
\end{equation}
and which require the presence of a second current of atoms propagating in the opposite direction, such that each atom first interacts with resonator 2 and then with resonator 1, see Fig.~\fref{fig:two_beams}. After imposing this condition, one derives an effective dissipative dynamics for the Bogoliubov modes $\hat b_1$ and $\hat b_2$, defined as in Eq.~(\ref{eq:boperators}), and which is valid over a time step $\Delta t$ such that there is at most one atom inside the resonator at a time. Provided that the atoms are initially prepared in the state $\ket +$ ($\ket -$) and the detuning is such that $\Delta=\Delta_0$ ($\Delta=-\Delta_0$), the master equation for the field density matrix in the coarse-grained time-scale is
\begin{eqnarray}
\pdiff{\hrho_t}{t}&=&r_\at\Delta \hrho^{\rightarrow}(t)~+r_\at\Delta \hrho^{\leftarrow}(t)\nonumber\\
&=& -\gamma \left[\bdag_j\hat b_j\hrho-2\hat
b_j\hrho\bdag_j+\hrho\bdag_j\hat b_j\right],
\label{dissipation_first_step}
\end{eqnarray}
where $\Delta\hrho^{\rightarrow}_t=\hrho^{\rightarrow}(t+\Delta t)-\hrho(t)$ (same for $^{\leftarrow}$), and $r_\at$ is the atom pump rate, which is assumed to be the same in both directions. Hence, mode $\hat b_1$ ($\hat b_2$) is exponentially damped according to an equation of the form given in Eq.~(\ref{master:eq}). Preparation of the two resonators in a two-mode squeezed state then is achieved, provided that both dynamics take place, by changing detuning and state preparation after the first step has been implemented after a sufficiently long time.

As in the protocol for entangling two modes of the same resonator, this scheme does not require detection of final atomic states, nor control of the atomic velocities. Moreover, the velocity of the atoms can change during propagation, without affecting the efficiency of the protocol. This result has been derived considering Hamiltonian dynamics between cavity and individual atoms.

Let us now discuss the limitations to this proposal. First, only one atom at a time must be present inside the cavities. Moreover, the atoms must not decay before they have interacted with both resonators. Atomic lifetime of the order of tens of milliseconds and atomic velocities of about 400 m/s require that the distance between the cavities is no larger than a few meters. Moreover, the resonators must be stable over the whole run of the experiment.

Experimental implementation of this proposal is quite challenging, since one must have a geometry with two counterpropagating atomic beams, thereby avoiding collisions between the atoms. As an alternative, one could think of a ring, as realized for instance in ion-storage setups~\cite{IonRing} in which the atomic beam is confined and which crosses two resonators placed at two different points of the ring. Another possibility is to implement an atomic fountain, where atoms traveling upwards and downwards interact with two vertically arranged cavities above the fountain~\cite{Atomic:Fountain}, or a optical conveyor belt~\cite{Belt}, where the atoms are transported back and forth between the resonators. These mechanisms have been employed so far in the optical regime, where the lifetime of the resonator modes is limited to tens of microseconds, in which case the implementation of this protocol is not straightforward.

\section{Conclusions and Outlook} \label{Sec:5}

In this paper we have extensively characterized the properties of an atomic beam as a reservoir for the modes of the electromagnetic field inside a resonator, for the purpose of creating entangled states of the cavity modes. The atoms can mediate the interaction between the modes of the same resonator or of distant cavities, establishing a dynamics whose steady state is an Einstein-Poldosky-Rosen entangled state. As opposed to previous proposals, see for instance~\cite{Wellens00,Law}, the atoms do not need to be initially correlated nor their number has to be controlled. Control on atomic velocity (interaction time) and atomic detection are not required. The degree of entanglement is controlled by an external maser field, which drives the atoms and tailors their interaction with the cavity modes. In this respect, the proposals discussed in this article are instances of quantum reservoir engineering. Statistical properties of the cavity field can be evaluated by measuring the internal states of the emerging atoms \cite{PhysRevA.49.2962}. Its state can also be determined by reconstructing the corresponding Wigner function, by suitably generalizing the schemes proposed in~\cite{lougovski-2003-91}.

The experimental setup, where these protocols could be implemented, is typical of microwave cavity quantum electrodynamics. Our proposal extends to cavity QED the technique of quantum reservoir engineering, originally applied to trapped ions. It offers a convenient way of implementing quantum mechanical dynamics and state preparation. Besides constituting a robust procedure for generating nonclassical states of the electromagnetic field in cavities in a steady-state regime, it might be useful for quantum networking with continuous variables in the microwave regime.

\begin{acknowledgments} Support by the European Commission (EMALI, MRTN-CT-2006-035369; SCALA, Contract No.\ 015714) and by the Spanish Ministerio de Innovaci\'on y Ciencia (Consolider Ingenio 2010 QOIT, CSD2006-00019; QNLP, FIS2007-66944; Ramon-y-Cajal program) are acknowledged. SP acknowledges support from the Studien\-stiftung des Deutschen Volkes. LD acknowledges support from the Brazilian agencies CNPq and FAPERJ and the National Institute of Science and Technology for Quantum Information. GM acknowledges support from the German Research Council (Heisenberg-professorship program).
\end{acknowledgments}


\begin{thebibliography}{10}
\bibitem{footnote}To be precise, $\gamma=rg^2\left<\tau^2\right>$, $<>$ denoting the average. Due to the narrowness of the distribution, however, we can write $\left<\tau^2\right>=\bar \tau^2$.
\end{thebibliography}
\end{document}